\documentclass[twocolumn,twocolappendix]{aastex701}
\usepackage{graphicx}
\usepackage{amsmath} 
\usepackage{amssymb} 
\usepackage{subcaption}
\usepackage{url}
\usepackage{enumitem}
\setlist[enumerate]{itemsep=2pt, parsep=0pt, topsep=4pt}
\begin{document}

\title{Investigating the initial conditions of globular clusters via $N$-body simulations: I. \\ half-mass radius and stellar density }

\author[orcid=0009-0007-3312-6039, gname=Ruoyun, sname=Huang]{Ruoyun Huang}
\affiliation{Department of Astronomy, School of Physics and Astronomy, Sun Yat-sen University, Zhuhai, Guangdong Province, China}
\affiliation{CSST Science Center for the Guangdong-Hong Kong-Macau Greater Bay Area, Zhuhai 519082, China}
\email{huangry8@mail2.sysu.edu.cn}


\author[orcid=0000-0002-0066-0346,gname=Baitian, sname=Tang]{Baitian Tang} 
\affiliation{Department of Astronomy, School of Physics and Astronomy, Sun Yat-sen University, Zhuhai, Guangdong Province, China}
\affiliation{CSST Science Center for the Guangdong-Hong Kong-Macau Greater Bay Area, Zhuhai 519082, China}
\email[show]{tangbt@mail.sysu.edu.cn}

\author[orcid=0000-0001-8713-0366,gname=Long, sname=Wang]{Long Wang} 
\affiliation{Department of Astronomy, School of Physics and Astronomy, Sun Yat-sen University, Zhuhai, Guangdong Province, China}
\affiliation{CSST Science Center for the Guangdong-Hong Kong-Macau Greater Bay Area, Zhuhai 519082, China}
\email[show]{wanglong8@sysu.edu.cn}

\author[orcid=0000-0002-1959-6946,gname=Holger, sname=Baumgardt]{Holger Baumgardt}
\affiliation{School of Mathematics and Physics, The University of Queensland, St. Lucia, QLD 4072, Australia}
\email{h.baumgardt@uq.edu.au}

\author[orcid=0000-0001-6820-1683,gname=Xia, sname=Li]{Xia Li}
\affiliation{Department of Astronomy, School of Physics and Astronomy, Sun Yat-sen University, Zhuhai, Guangdong Province, China}
\affiliation{CSST Science Center for the Guangdong-Hong Kong-Macau Greater Bay Area, Zhuhai 519082, China}
\email{lixia76@mail2.sysu.edu.cn}

\author[orcid=0009-0007-4049-127X,gname=Genhao, sname=Liu]{Genhao Liu}
\affiliation{Department of Astronomy, School of Physics and Astronomy, Sun Yat-sen University, Zhuhai, Guangdong Province, China}
\affiliation{CSST Science Center for the Guangdong-Hong Kong-Macau Greater Bay Area, Zhuhai 519082, China}
\email{liugh33@mail2.sysu.edu.cn}

\author[orcid=0000-0001-6590-8122,gname=Jundan, sname=Nie]{Jundan Nie}
\affiliation{CAS Key Laboratory of Optical Astronomy, National Astronomical Observatories, Chinese Academy of Sciences, Beijing 100101, People’s Republic of China}
\email{jdnie@nao.cas.cn}

\author[orcid=0000-0003-3347-7596,gname=hao, sname=Tian]{Hao Tian}
\affiliation{Key Laboratory of Space Astronomy and Technology, National Astronomical Observatories, Chinese Academy of Sciences, Beijing 100101, People’s Republic of China}
\email{tianhao@nao.cas.cn}

\begin{abstract}

Born out of the turbulent, dense gas of the early Universe, globular clusters (GCs) represent a significantly different star-formation mode compared to the local Universe. To quantify the initial cluster conditions, we perform the long-term dynamical evolution of GCs using high‑precision direct $N$-body simulation code {\sc petar}. As the first step in a series of studies, 
we investigate the two low‑mass outer‑halo GCs, Eridanus and Whiting~1.  By comparing simulated number density profiles with observations, we derive best‑fit initial half‑mass radii of $2.0 \pm 1.0$ pc for Eridanus and $2.5 \pm 0.5$ pc for Whiting~1, corresponding to initial surface densities of $2592.03\ M_\odot \cdot \text{pc}^{-2}$ and $226.35\ M_\odot \cdot \text{pc}^{-2}$, respectively --- $2 - 3$ orders of magnitude higher than their present‑day values. The absence of multiple populations (MPs) in Whiting~1 and young massive clusters (YMCs), in contrast to the MP‑hosting Eridanus, supports the scenario that MP formation requires higher initial surface density and/or lower metallicity. Using the simulations as a template, we identify five high‑probability extra‑tidal candidates of Whiting~1 from Gaia DR3 and APOGEE DR17, which exhibit coherent kinematics and chemical abundances. Our results demonstrate that the diffuse states of low‑mass GCs are the product of long‑term dynamical expansion and tidal stripping, and that their primordial configurations were significantly denser.

\end{abstract}



\section{Introduction} 
Globular clusters (GCs) are among the oldest and most massive bound stellar systems in the Universe. As relics of the early epochs of galaxy formation, they serve as crucial tracers for reconstructing the assembly history of the Milky Way (MW). Distributed throughout the Galactic halo and bulge, these dense systems provide unique insights into the physical conditions of the primitive environment in which they formed\citep{brodie2006extragalactic}.

Intense star formation in the chaotic early universe gave rise to distinctive element patterns rarely seen in present-day star-forming regions.
Contrary to the traditional view of GCs as simple stellar populations (SSP) with stars of the same age and chemistry, a growing number of studies reveal they host multiple stellar populations (MPs, \citealt{gratton2004abundance,milone2022}).
Star clusters hosting MPs display significant star-to-star variations in their chemical abundances. Generally, stars with chemical compositions similar to those of field stars are defined as the primordial generation (1G). In contrast, the second generation (2G) is characterized by distinct chemical anomalies, specifically showing enrichments in N, Na, and Al, coupled with a depletion of C, O and Mg \citep{carretta2009anticorrelationa,carretta2009anticorrelationb,pancino2017gaia,tang2017}.

To integrate these chemical enrichment patterns into a unified formation framework, researchers have sought to identify the fundamental physical parameters that define the essence of MPs/GCs, such as cluster mass \citep{carretta2010properties, tang2021, wangMultiplePopulationsLowmass2023b}, age \citep{Bastian2018}, metallicity, and compactness \citep{krauseGasExpulsionMassive2016,huang2024driving}. However, this endeavor faces significant challenges. Having undergone dynamical evolution over timescales of several gigayears (Gyr), GCs have suffered severe mass loss and structural transformation under the combined effects of two-body relaxation, stellar evolution, and Galactic tidal stripping. These dynamical processes have largely ``erased'' the memory of the clusters' initial conditions, making it difficult to accurately retrieve their primordial physical parameters based solely on present-day observational properties. Consequently, this obscures the fundamental boundaries in formation mechanisms between MP/GCs and SSP/OCs (open clusters).

Leveraging the high sensitivity of JWST/NIRCam and the power of gravitational lensing, researchers have identified populations of star clusters (or clumps) within high-redshift galaxies. Notable examples include candidates in the Sparkler galaxy (cluster masses $> 10^6 M_\odot$, \citealt{mowlaSparklerEvolvedHighredshift2022}), the Cosmic Gems arc ($\sim 10^6 M_\odot$, \citealt{adamoBoundStarClusters2024c}), and the SMACS 0723 field ($10^5 \text{--} 10^9 M_\odot$, \citealt{10.1093/mnras/stac3791}). These discoveries provide pivotal clues for understanding the formation and early dynamical evolution of GCs during the epoch of reionization.

Galactic GCs currently span a broad mass range ($10^3 - 10^6 M_\odot$, \citealt{baumgardt2018catalogue} ). Although JWST has successfully detected the most massive high-redshift clusters via gravitational lensing, infant low-mass clusters ($< 10^5 M_\odot$) remain largely inaccessible due to current detection limits. This observational gap hinders the complete reconstruction of the cluster mass function during the epoch of reionization, complicating efforts to physically link distant high-redshift compact objects with evolved Galactic clusters. 
This challenge is particularly pronounced when constraining key structural parameters such as the initial radius, for which direct observation is currently virtually impossible.

Consequently, employing numerical simulations to reconstruct the dynamical history of star clusters within the Galactic potential has become key to bridging this gap. In this study, we employ the high-precision $N$-body dynamics code {\sc petar}\citep{wangPeTarHighperformanceNbody2020} to analyze a sample of approximately 20 clusters situated at the mass transition region between OCs and GCs (with present-day masses of $10^3 - 10^5 M_\odot$), with the aim of investigating their initial conditions. As the first step in a series of studies, this paper focuses on Eridanus and Whiting~1—two representative relatively low-mass clusters. By reproducing their evolutionary paths, we verify the feasibility and accuracy of our numerical model in resolving evolutionary processes within this specific mass regime.

Eridanus is a low-mass GC located in the distant outer MW halo. It has a total mass of approximately $9.3 \times 10^3 M_\odot$ and a heliocentric distance of 90.10 kpc\citep{baumgardt2019mean}. Despite its low mass, spectroscopic studies have unexpectedly confirmed the presence of MPs within the cluster \citep{wangMultiplePopulationsLowmass2023b}. Compared to other GCs at similar Galactocentric distances, Eridanus exhibits a significantly higher metallicity ([Fe/H]$\approx -1.45$, \citealt{harris2010new}); meanwhile, its estimated age of $10-10.5$ Gyr\citep{beccari2012central,wangMultiplePopulationsLowmass2023b} is slightly younger than that of inner-halo clusters with comparable metallicities, suggesting it may possess a unique accretion history. Furthermore, the exceptionally extended and diffuse spatial structure of this cluster, along with its accompanying tidal stream features \citep{myeongTidalTailsOuter2017a}, provides clear observational evidence that it is currently undergoing active dynamical disruption.

Whiting~1 is a faint and young GC within the Galactic halo, characterized by an age of $5-6.5$ Gyr\citep{carraroWhitingNewHalo2005,carraroWhitingYoungestGlobular2007} and a stellar mass of only $\sim 2,000 M_\odot$\citep{baumgardtStellarMassSegregation2022a} . 
It is widely considered to be associated with the Sagittarius dwarf spheroidal galaxy (Sgr dSph), representing a remnant of an accreted system. \citet{nie2022searching} identified tidal tails extending from both sides of the cluster, which are precisely aligned with the orbital trajectories of both Whiting~1 and the Sgr dSph. Despite its dynamic history, recent spectroscopic analysis indicates that Whiting~1 remains chemically homogeneous and does not exhibit MPs \citep{huang2024driving}.

\begin{deluxetable*}{ccccccccc}
\tabletypesize{\footnotesize} 
\tablewidth{0pt}
\tablecaption{The observational parameters of Eridanus and Whiting~1. \label{tab:description}}
\tablehead{
\colhead{Object} & \colhead{RA} & \colhead{DEC} & \colhead{$\mu_\alpha^*$} & \colhead{$\mu_\delta$} & \colhead{RV} & \colhead{$R_\odot$} & \colhead{Age} & \colhead{[Fe/H]} \\
\colhead{} & \colhead{(deg)} & \colhead{(deg)} & \colhead{(mas/yr)} & \colhead{(mas/yr)} & \colhead{(km/s)} & \colhead{(kpc)} & \colhead{(Gyr)} & \colhead{(dex)} \\
}
\startdata
Eridanus  & 66.18561 & -21.18678 &  0.477 & -0.290 &  -22.91 & 84.68 & 10 & -1.43 \\
Whiting~1 & 30.73750 &  -3.25278 & -0.244 & -2.019 & -130.41 & 30.59 &  5 & -0.70 \\
\enddata

\tablecomments{Structural and kinematic parameters for Eridanus and Whiting~1 are retrieved from the Galactic Globular Clusters Database (v4, 2023 March; \url{https://people.smp.uq.edu.au/HolgerBaumgardt/globular/}). Distances and mean motions are based on \citet{vasiliev2021gaia} and \citet{baumgardt2021accurate}. Ages are adopted from \citet{baumgardt2019mean}, and metallicities are taken from the Harris catalog \citep{harris2010new}.}

\end{deluxetable*}

This paper is organized as follows:
Section 2 details the numerical setup of our $N$-body simulations.
Section 3 cross-compares the observational data with our simulation results to deduce the initial parameters of Eridanus and Whiting~1, and explores their dynamical evolution.
In Section 4, by comparing the current cluster parameters with the simulation results, we analyze the overall evolutionary trends of GCs within density space. Furthermore, leveraging the phase-space information provided by the simulations, we identify potential escaped members of Whiting~1 and predict their observational signatures in the Galactic halo.
Finally, Section 5 summarizes our core conclusions and provides concluding remarks.

\section{$N$-body simulation} \label{sec:style}

\subsection{{\sc petar}}

Direct $N$-body simulations represent the most accurate method for studying the dynamical evolution of stellar systems; however, their computational cost increases steeply with the number of particles. To overcome this bottleneck, in this work, we employ the new-generation high-performance $N$-body code, {\sc petar}, to model our target clusters.

{\sc petar} is specifically designed to simulate gravitationally bound systems containing numerous hierarchical subsystems (e.g., binaries and triples) and achieves efficient parallelization through the FDPS framework\citep{wangPeTarHighperformanceNbody2020,oshino2011particle,iwasawa2017pentacle,iwasawa2016implementation,iwasawa2020accelerated}. The code employs a hybrid integration strategy to balance accuracy and efficiency: long-range gravitational forces are computed using the Barnes–Hut tree algorithm, which significantly reduces complexity for large-scale systems, while short-range strong interactions are handled by a fourth-order Hermite scheme combined with the Slow-Down Algorithmic Regularization (SDAR) method to ensure precise tracking of compact multiple systems\citep{barnes1986hierarchical,wang2020slow}. 

Given this superior computational efficiency, we utilize {\sc petar} to perform full dynamical simulations—incorporating stellar evolution—of two low-mass GCs, Whiting~1 and Eridanus. Our goal is to investigate their possible initial configurations, compare the simulated tidal streams with observations, and further explore the origin and evolutionary history of these clusters.

Within the dynamical evolution framework of this study, our core methodology is to constrain the initial structure of the clusters—specifically, their initial half-mass radii ($r_{h,0}$)—using {\sc petar} $N$-body simulations. The rationale for prioritizing $r_{h,0}$ as the primary initial parameter resides in the fact that it directly characterizes the macroscopic spatial scale and the depth of the cluster's gravitational potential well, and it can be precisely configured in the {\sc petar} initialization module. Physically, $r_{h,0}$ determines the initial compactness of the cluster, thereby strictly setting the baseline scale for the two-body relaxation time ($t_{rh}$)\citep{baumgardt2003dynamical}. Specifically, a smaller $r_{h,0}$ endows the cluster with an extremely short relaxation time, significantly accelerating the processes of mass segregation and core collapse. Conversely, a larger $r_{h,0}$ results in a looser cluster structure, rendering it more susceptible to external tidal stripping, which in turn accelerates stellar escape and cluster dissolution.

Because subtle variations in $r_{h,0}$ are amplified over long-term dynamical evolution and ultimately imprint distinct signatures onto the present-day surface number density profiles, we can establish a robust mapping from initial states to current observations. 

Given that directly quantifying the 3D half-mass radius ($r_h$) during both simulation post-processing and actual observations introduces significant systematic errors and technical challenges, this study adopts the 2D surface number density profile as the core metric to constrain $r_{h,0}$\citep{miocchiSTARCOUNTDENSITY2013}. A high degree of statistical agreement between the simulated density profiles and the observational data will serve as conclusive validation for the physical reliability of our adopted $r_{h,0}$. For both clusters, we adopted  $r_{h,0} = 2, 3,$ and $4$ pc.

For comparison with our simulations, we selected the observational data as follows. For the Eridanus cluster, we adopted the stellar number density profile compiled by \citet{baumgardt2023evidence}. This dataset comprises high-precision inner photometry obtained from the Hubble Space Telescope (HST; WFPC2 for the inner cluster parts) with observations of the outer region from CFHT \citep{munoz2018megacam}. Detailed procedures for the cross-calibration, spatial stitching, and artificial star tests of these two datasets are comprehensively described in \citet{baumgardt2023evidence}.
For the number density profile of Whiting~1, \citet{carraroWhitingYoungestGlobular2007} obtained deep $BVI$ CCD imaging using the FORS1 instrument on the Very Large Telescope (VLT).
These observations reach a limiting magnitude of $V \approx 27$ mag, allowing for a clear resolution of the subgiant branch (SGB), the main-sequence turnoff (MSTO), and the unevolved main sequence extending roughly three magnitudes below the turnoff. It significantly improves the quality and resolution of the cluster’s color–magnitude diagram (CMD).

\subsection{Initial condition}

To determine the initial position and velocity of the clusters, we performed a backward orbit integration using a single test particle representing the cluster's center of mass. The integration was initialized with present-day 6D phase-space coordinates derived from observational data and evolved backward within the MWPotential2014 galactic potential \citep{bovy2015galpy} for a duration corresponding to the cluster's estimated age (Table \ref{tab:description}). This procedure yielded the initial coordinates $(x_0, y_0, z_0)$ and velocities $(v_{x0}, v_{y0}, v_{z0})$ at the birth epoch. We validated these initial conditions by performing a forward evolution to ensure the final state closely matches the observed phase-space distribution.

To generate the initial conditions for the simulated star clusters, we utilized the McLuster code \citep{kupper2011mass}, assuming that the initial spatial density distribution of the clusters follows a Plummer model. Guided by the numerical results of \citet{baumgardt2019mean} and  \citet{baumgardt2023evidence}, we set the initial total masses of Eridanus and Whiting~1 to $3.66 \times 10^4\,M_\odot$ and $5.0 \times 10^3\,M_\odot$, respectively.  For the internal stellar populations, we adopted the ``bottom-light'' initial mass function (IMF) proposed by \citet{baumgardt2023evidence}. Recent studies based on a large sample of GCs indicate a universal depletion of low-mass stars; therefore, compared to the canonical Kroupa IMF, this adopted IMF is flatter at the low-mass end.
The adopted IMF is defined as the following three-part power law:

\begin{align}
  \alpha_{\text{low}}  &= -0.3 \pm 0.2 \quad (m < 0.4\,M_\odot) \\
  \alpha_{\text{med}}  &= -1.65 \pm 0.20 \quad (0.4 < m < 1.0\,M_\odot) \\
  \alpha_{\text{high}} &= -2.30 \pm 0.20 \quad (m > 1.0\,M_\odot)
\end{align}

Under this distribution, the generated stellar masses were restricted to the range of $0.08\,M_\odot \leq M \leq 150\,M_\odot$.  Accordingly, we generated initial single-star samples of approximately 30,000 and 5,000 stars for the two clusters, with their metallicities set to $Z=0.0007$ and $Z=0.0041$, respectively. Considering the computational cost of the evolutionary calculations, we neglected binaries in the initial models. Finally, we evolved the two samples to ages of 10 Gyr and 5 Gyr, respectively, using the \texttt{petar.bse} module. This module employs a significantly updated BSE package \citep{hurley2000comprehensive, hurley2002evolution} and incorporates state-of-the-art stellar evolutionary recipes to achieve self-consistent stellar evolution.

Finally, the photometric properties of the evolved stars were determined using the {\sc galevnb} code \citep{pang2016galevnb}. Based on the fundamental stellar parameters derived from the $N$-body simulations---such as stellar mass, effective temperature, luminosity, and metallicity---this code selects the corresponding spectral templates from the stellar library of \cite{lejeune1997standard,lejeune1998standard}. These synthetic spectra are then convolved with the response curves of the designated filters to yield the mock magnitudes. To facilitate a direct comparison with observational data, the simulated magnitudes were ultimately transformed into the standard Johnson photometric system ($V$ and $I$ bands).

\section{Results} \label{sec:floats}

\subsection{Initial half-mass radii}

To ensure a consistent comparison with actual observations, we post-processed the final states of the $N$-body simulations to extract the 2D surface number density profiles. When deriving the number density distribution of the simulated clusters, we excluded neutron stars, black holes, and white dwarfs from our models. Regarding the formation of these compact objects, we adopted a Hobbs distribution with a velocity dispersion of $\sigma = 265$ km/s for their natal kicks. While these objects drive the dynamical evolution of the cluster within the simulation, they were not included in the final number density calculation due to their invisibility in optical observations.

Owing to its substantial initial mass and consequently deeper potential well, Eridanus exhibits only marginal sensitivity in its final structural parameters to the $r_{h,0}$ adopted. This suggests that the cluster is relatively resilient to tidal stripping and internal dynamical processes. In contrast, the lower-mass Whiting~1 shows a much stronger dependence on its initial configuration: a larger $r_{h,0}$ significantly increases its susceptibility to tidal mass loss, leading to a more pronounced stellar mass loss under the combined effects of the Galactic tidal field and internal relaxation, ultimately resulting in a highly diffuse final structure.

To ensure consistency between observations and simulations, we applied uniform spatial and magnitude cuts to the stellar samples. In terms of spatial sampling, the boundary for Eridanus was set to its tidal radius (383.5 arcsec), whereas for Whiting 1, limited by the coverage of the actual observational field of view, the boundary was set to 200 arcsec \citep{webb2013influence,vasiliev2021gaia,baumgardt2018catalogue}. Regarding the magnitude selection, both the observational and simulated data were uniformly restricted to a secure range of $V \le 24.0$ mag \citep{baumgardt2023evidence,carraroWhitingYoungestGlobular2007}.

\begin{figure}[ht]
    \centering
    
    \begin{subfigure}{\linewidth}
        \centering
        \includegraphics[scale=0.3]{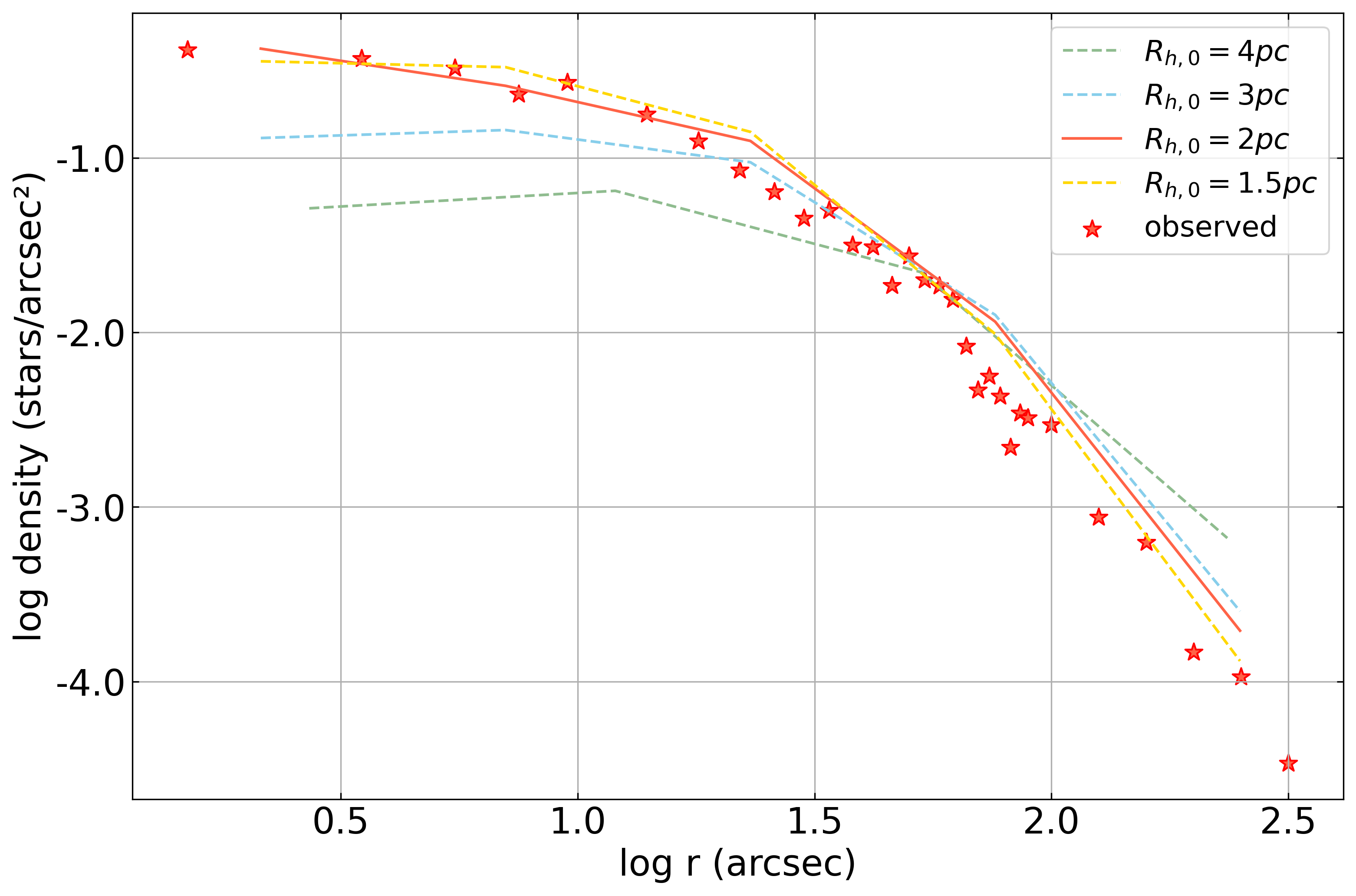}
        \caption{Eridanus}
        \label{fig:ndEr}
    \end{subfigure}
    
    \vspace{1em} 
    
    \begin{subfigure}{\linewidth}
        \centering
        \includegraphics[scale=0.3]{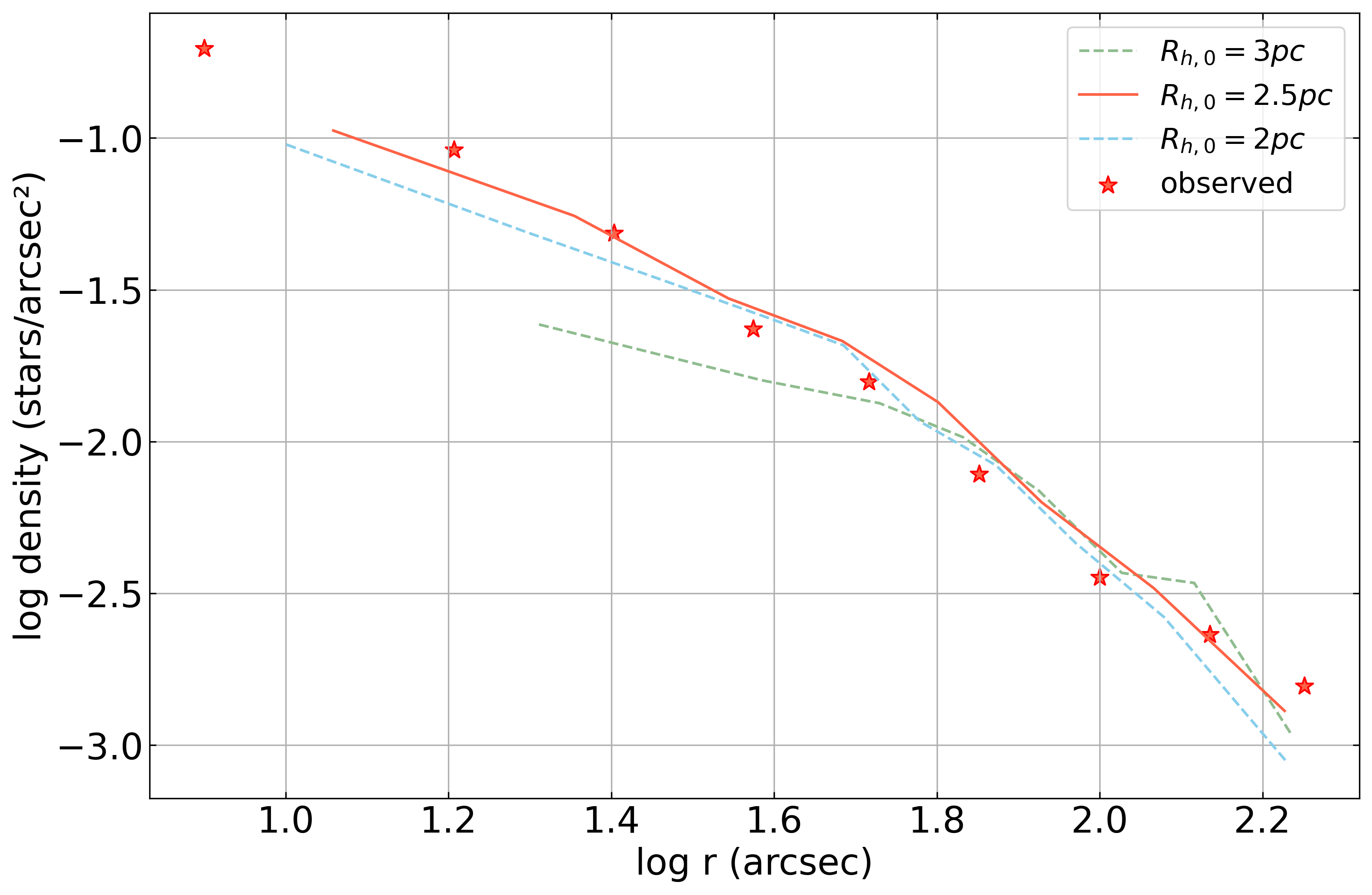}
        \caption{Whiting~1}
        \label{fig:ndw1}
    \end{subfigure}
    
    \caption{Number density distributions of the GCs Eridanus (a) and Whiting~1 (b). The solid lines represent the simulation results for different half-mass radii, while the red star symbols denote the observational data.}
    \label{fig:nd_combined}
\end{figure}

By comparing the simulated number density profiles with the observational data (Figure \ref{fig:ndEr}), the best-fit $r_{h,0}$ is found to be 2 pc for Eridanus, with an uncertainty of 1 pc. This preferred initial radius is further supported by its stellar mass function and mass segregation signature across the inner and outer regions (see Appendix~\ref{app:mass_function}).For Whiting~1 (Figure \ref{fig:ndw1}), the observed number density profile falls between the predictions of the models initialized with$r_{h}$ of $2\text{ pc}$ and $3\text{ pc}$. Consequently, we performed a supplementary simulation with $r_{h,0} = 2.5\text{ pc}$, which ultimately yielded the best agreement with the observational data. Therefore, the initial $r_{h,0}$ is found to be $ 2.5 \pm 0.5 \text{ pc}$.
All subsequent simulated data plots in this paper are generated based on the respective best-fitting half-mass $r_{h,0} $ of the two clusters (2 pc for Eridanus and 2.5 pc for Whiting~1).

The present-day central number density of the star cluster does not exhibit a simple monotonic relationship with its $r_{h,0}$. Our simulations verify that for the Whiting~1, when $r_{h,0}$ is set to 1.5 pc, the extremely short relaxation time of the system triggers early core collapse, inevitably leading to the complete dissolution of the cluster before it reaches its present age. At the present-day epoch, the initially looser 2.5 pc model counter-intuitively possesses a higher central number density than the initially more compact 2.0 pc model. This occurs because the 2.0 pc model evolves more rapidly, whereas the 2.5 pc model benefits from a longer relaxation timescale, thereby retaining more core mass.

Our simulations indicate that to reproduce the observed present-day $r_h$ of $18.83$ pc for Eridanus and $16.53$ pc for Whiting~1, the clusters must have originated from much more compact initial states ($r_{h,0} \approx 2.0 - 2.5$ pc). This implies a significant spatial expansion by a factor of $\sim 7-9$ throughout their evolutionary history. This substantial expansion is driven by the combined effects of stellar mass loss and dynamical relaxation. The mass loss due to stellar evolution in the early phase induces an adiabatic expansion of the cluster potential. Subsequently, two-body relaxation transports energy from the core to the outer regions, further extending the stellar halo.

Based on present-day observational constraints, Eridanus and Whiting~1 exhibit remarkably diffuse structures. Their 3D $r_h$ reach $18.83\text{ pc}$ and $16.53\text{ pc}$, corresponding to exceptionally low derived surface densities ($\Sigma$) of merely $7.41\ M_\odot \cdot \text{pc}^{-2}$ and $1.42\ M_\odot \cdot \text{pc}^{-2}$, respectively. As illustrated in Figure \ref{fig:den} (marked with crosses), when compared to other Galactic GCs (green dots)\footnote{https://people.smp.uq.edu.au/HolgerBaumgardt/globular/parameter.html
}, these two objects fall distinctly into the parameter space defined by low $\Sigma$ and extended radii, characterizing them as typical ``loose" or dynamically evolved clusters.

However, our numerical simulations reveal a starkly contrasting initial state: the best-fit models indicate $r_{h,0}$ of only $2.0\text{ pc}$ and $2.5\text{ pc}$, with corresponding initial surface densities as high as $2592.03$ and $226.35\ M_\odot \cdot \text{pc}^{-2}$(marked with stars). These values represent an enhancement of 2–3 orders of magnitude compared to their present-day conditions. This profound disparity in physical properties demonstrates that, over billions of years of evolution, the combined effects of internal dynamical relaxation and the external Galactic tidal field have driven substantial structural expansion and the continuous stripping of outer stars. Consequently, the observed low surface density is not an intrinsic primordial feature but rather a product of dynamical processing; this further implies that low-mass GCs, which currently appear extremely diffuse or near dissolution, likely possessed very high concentrations at the time of their formation.

\subsection{Structural and Dynamical Evolution}

The long-term evolution of a star cluster is governed by a combination of internal and external processes. The internal mechanisms are primarily stellar evolution and two-body relaxation. Stellar evolution itself initially drives the mass loss of the system. Concurrently, two-body relaxation induces significant mass segregation: massive stars gradually sink toward the cluster center due to the loss of kinetic energy, while low-mass stars gain kinetic energy through energy exchange, are pushed to the cluster outskirts, and ultimately escape. This dynamical evaporation process further exacerbates the mass loss of the cluster. Driven jointly by the aforementioned mechanisms, the total mass of the cluster continuously decays over time, while its outer Lagrangian radii exhibit a structural characteristic of continuous expansion.

Among external mechanisms, tidal shocks exert a crucial impact on the evolution of star clusters. For GCs within the MW, disc passages \citep{gnedin1997destruction, kruijssen2009dissolution, webb2014effects} and perigalactic passes \citep{gnedin1997destruction, baumgardt2003dynamical, webb2013influence} significantly impact their dynamical evolution \citep{gnedin1999effects}.

\begin{figure}[ht]
    \centering

    \begin{subfigure}{\linewidth}
        \centering
        \includegraphics[scale=0.32]{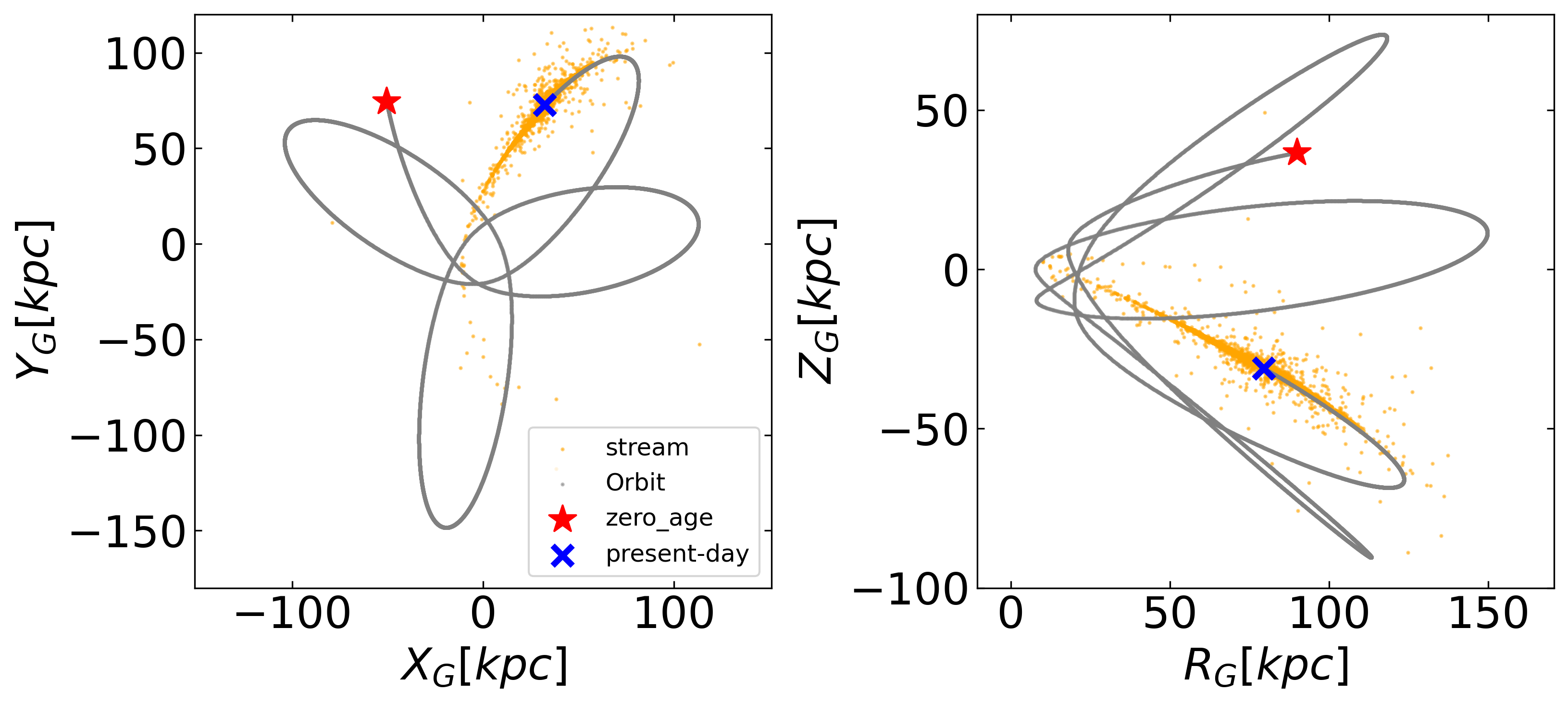}
        \caption{Eridanus}
        \label{fig:orbitEr}
    \end{subfigure}

    \vspace{1em} 

    \begin{subfigure}{\linewidth}
        \centering
        \includegraphics[scale=0.32]{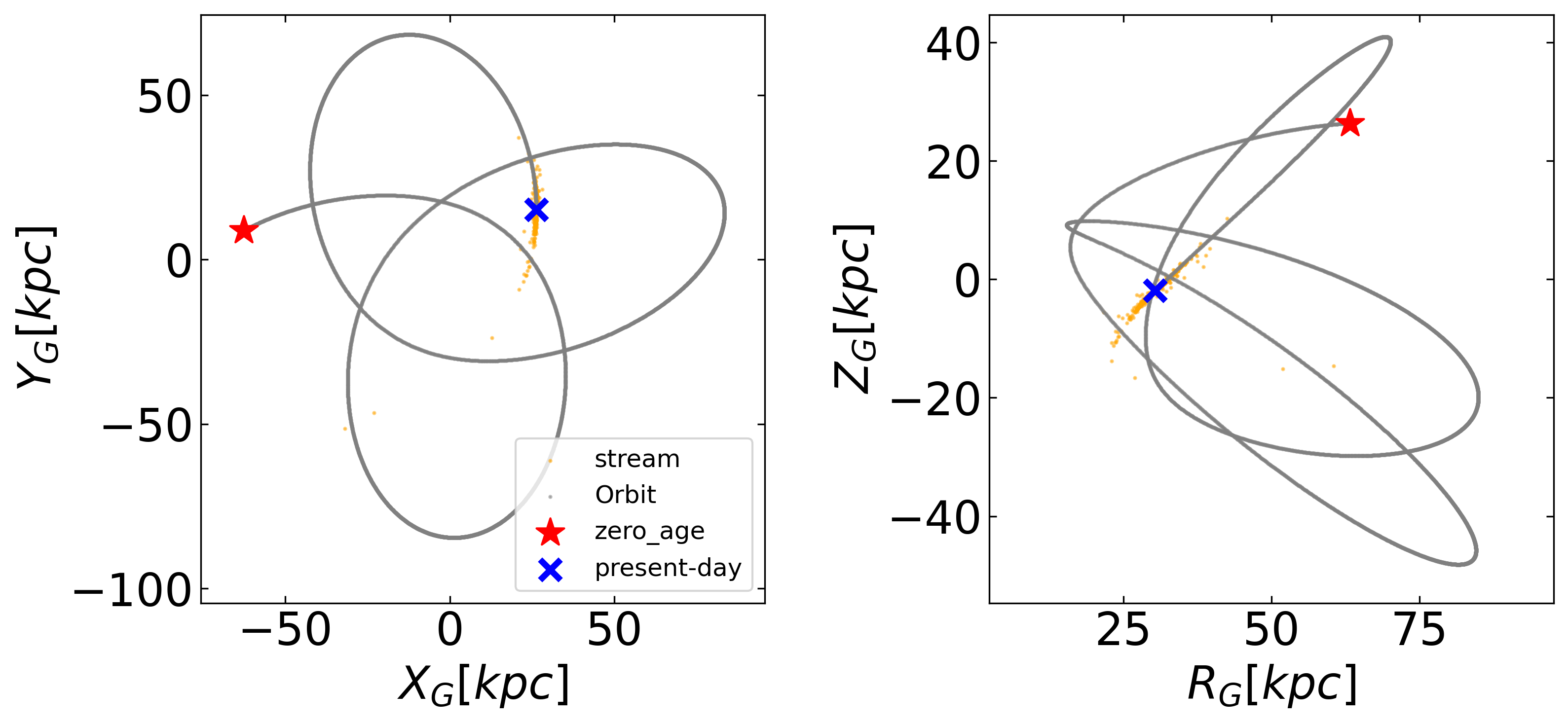} 
        \caption{Whiting~1}
        \label{fig:orbitw1}
    \end{subfigure}

    \caption{Orbits of (a) Eridanus and (b) Whiting~1 simulated with {\sc petar} (projected in the $X_G-Y_G$ and $R_G-Z_G$ planes). The grey solid line represents the trajectory, while the red star and blue cross denote the initial and current positions, respectively. Yellow dots indicate the stellar stream in the simulation.}
    \label{fig:orbits_combined}
\end{figure}

\begin{figure}[ht]
    \centering

    \begin{subfigure}{\linewidth}
        \centering
        \includegraphics[scale=0.3]{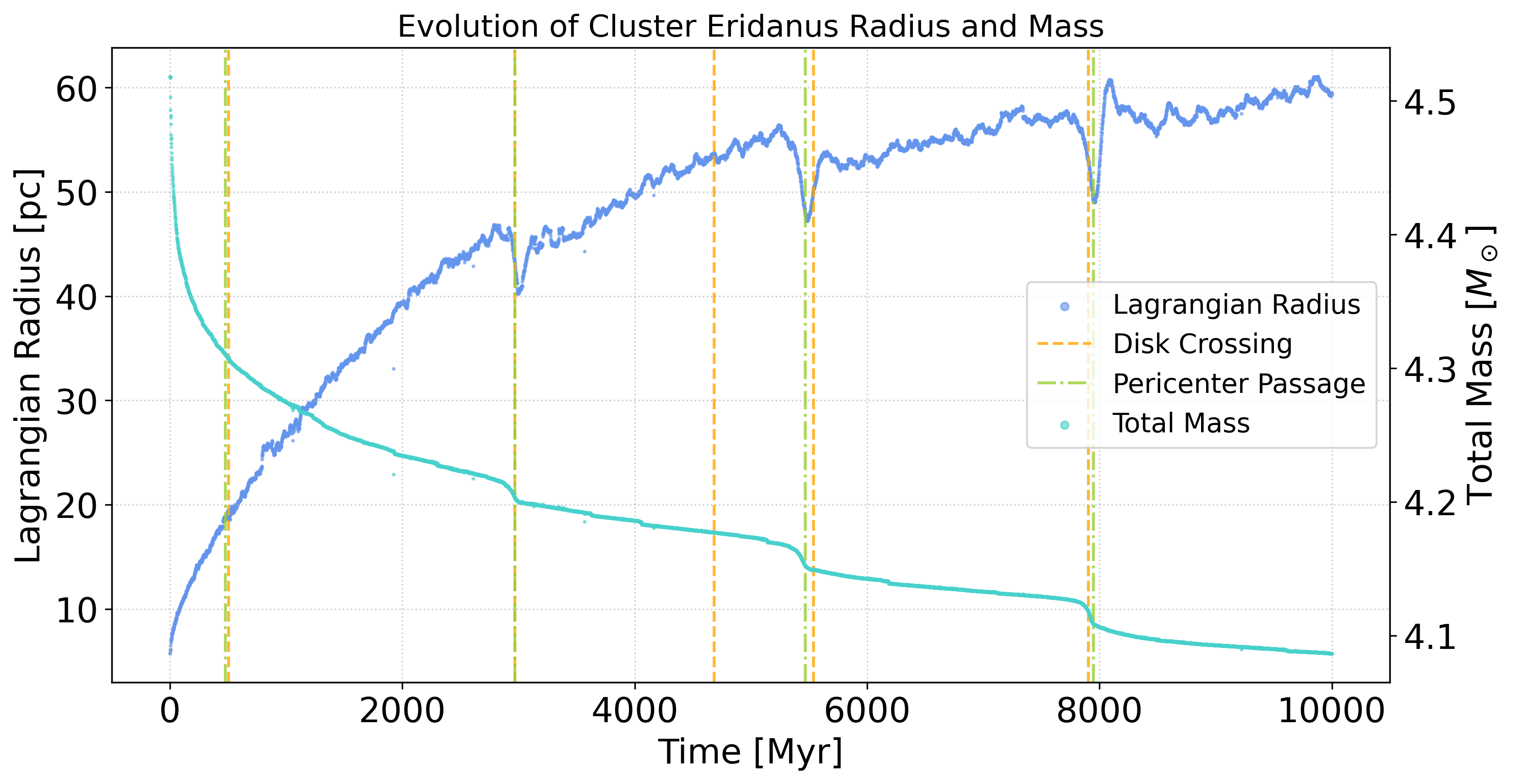}
        \caption{Eridanus}
        \label{fig:RL_mass_Er}
    \end{subfigure}

    \vspace{1em} 

    \begin{subfigure}{\linewidth}
        \centering
        \includegraphics[scale=0.3]{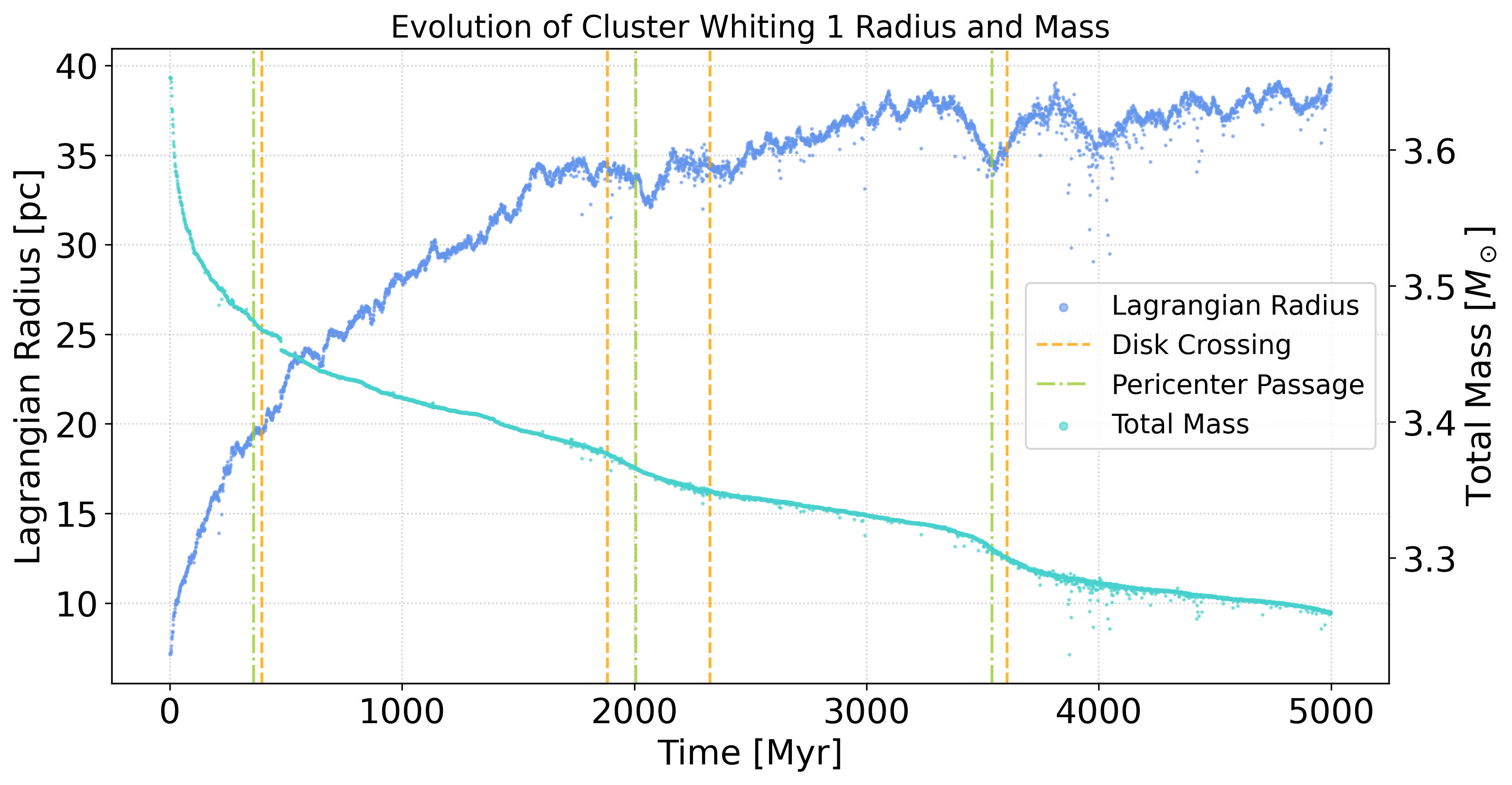}
        \caption{Whiting~1}
        \label{fig:RL_mass_w1}
    \end{subfigure}

    \caption{Evolution of the Lagrange radii (blue) and mass (cyan) of the GCs (a) Eridanus and (b) Whiting~1 from {\sc petar} simulations. The orange and green dashed lines indicate the epochs of disk crossing and pericentric passage, respectively. Significant variations in both radii and mass are observed when these two events occur in close succession.}
    \label{fig:RL_mass_combined}
\end{figure}

Based on our simulation results, we reconstructed the orbital evolutionary paths of the Eridanus and Whiting~1 clusters since their formation (illustrated by the solid grey lines in Figures \ref{fig:orbitEr} and \ref{fig:orbitw1}, respectively), where the yellow scatter points denote the spatial distribution of the tidal streams surviving to the present day. In the Galactocentric coordinate system adopted herein, $X_G$, $Y_G$, and $Z_G$ denote the 3D Galactocentric Cartesian positions projected within and perpendicular to the Galactic plane, respectively.By identifying the nodes where the orbital height $Z_G = 0$ and the minima of the galactocentric distance $R_G$, we precisely determined the critical dynamical moments of the clusters' disc passages and perigalactic passes (as indicated by the orange and green dashed lines in Figures \ref{fig:RL_mass_Er} and \ref{fig:RL_mass_w1}).

From a physical perspective, when a perigalactic pass highly coincides with a disc passage, the cluster typically suffers the most intense composite tidal perturbations, leading to noticeable episodic drops in both its Lagrangian radii and instantaneous mass at those specific moments. However, according to the study by \citealt{vesperiniEffectsDynamicalEvolution1997}, in the outer Galactic regions (large galactocentric distances, typically defined as $R_G > 10 - 15\text{ kpc}$), the matter distribution of the Galactic disc is extremely tenuous, and star clusters experience almost no significant gravitational perturbations during disc crossings; hence, the "disc shocking" effect can be safely neglected in such environments. The simulation results in this study further substantiate this theoretical expectation: since the galactocentric distances of both clusters in our models remain greater than 10 kpc even at their perigalacticons, the impulsive mass loss rates induced by a single disc passage remain relatively low (below 6.3\% in our simulations),, even during the aforementioned most drastic overlapping evolutionary phase. This indicates that external tidal shocks do not exert a destructive impact on their overall structures, and the long-term mass loss of both clusters is still predominantly governed by the internal process of two-body relaxation.

\begin{figure}[ht]
    \centering

    \begin{subfigure}{\linewidth}
        \centering
        
        \includegraphics[scale=0.5]{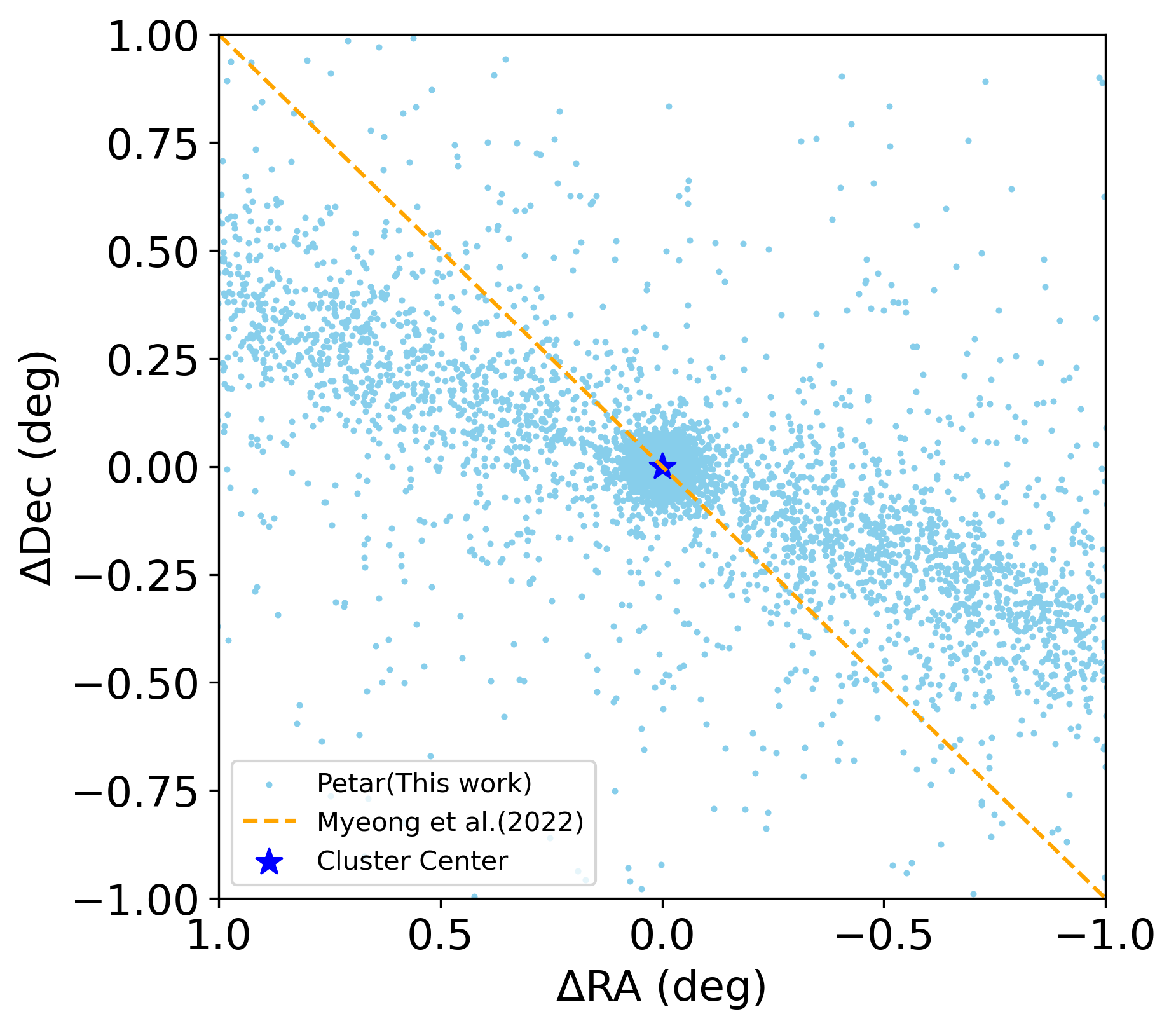}
        \caption{Eridanus stream}
        \label{fig:streamEr}
    \end{subfigure}

    \vspace{1em} 

    \begin{subfigure}{\linewidth}
        \centering
        \includegraphics[scale=0.5]{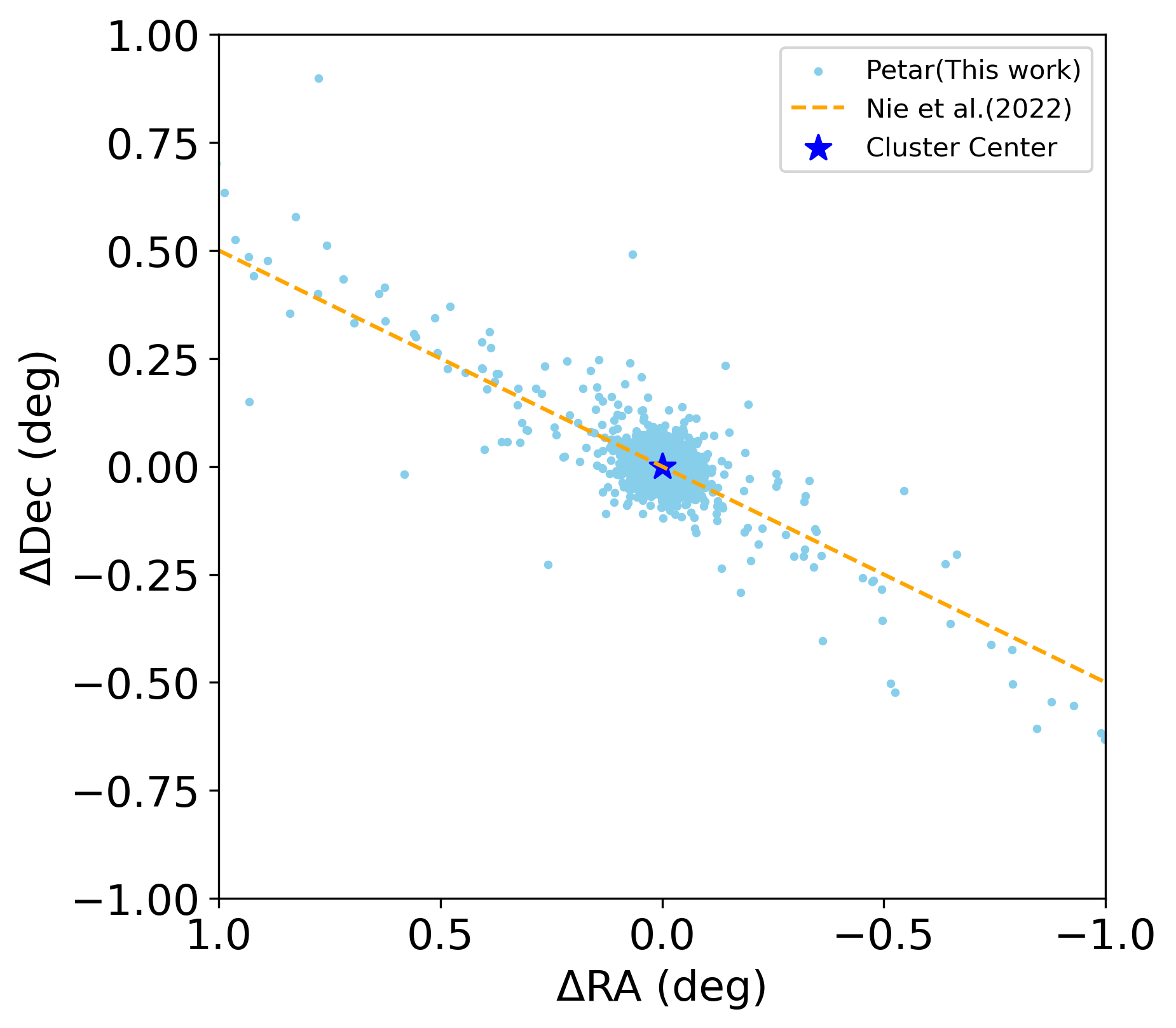}
        
        \caption{Whiting~1 stream}
        \label{fig:streamw1}
    \end{subfigure}

    \caption{Distribution in $\Delta\text{RA}$ and $\Delta\text{Dec}$ for (a) Eridanus and (b) Whiting~1 cluster streams. Orange dashed lines indicate the observed tracks of stellar streams.}
    \label{fig:streams_combined}
\end{figure}

Figures \ref{fig:streamEr} and \ref{fig:streamw1} present the two-dimensional spatial distributions of the Eridanus and Whiting~1 stellar streams generated by the {\sc petar} simulations (blue dots), respectively, superimposed with the directions of the tidal structures revealed by observations (orange solid lines). For the observational data, the tidal tail structure of Eridanus is extracted from the study by \citet{myeongTidalTailsOuter2017a} based on CFHT data, while that of Whiting~1 is extracted from the observational results of \citet{nie2022searching} based on deep DESI Legacy Imaging Surveys. The coordinate system for all panels adopts relative right ascension versus relative declination ($\Delta\text{RA}$ vs. $\Delta\text{Dec}$). A comparison indicates that the simulated stellar streams and the actual observed tidal tails exhibit a high degree of consistency in their overall spatial extension directions. 

In the $N$-body simulations of this study, we adopted an exclusively static Galactic background gravitational potential (MWPotential2014), without incorporating the gravitational perturbations from nearby dwarf galaxies (such as the Large Magellanic Cloud; LMC) or accounting for the dynamical evolution of the Milky Way potential itself. Particularly for the dynamical evolution of Whiting~1, the gravitational potential of the Sagittarius (Sgr) dSph galaxy was omitted. Remarkably, however, the simulated extra-tidal structures of Whiting~1 not only highly coincide with the observed morphology but also exhibit a natural and heavy spatial alignment with the large-scale Sgr stellar stream in 3D space. This striking dynamical synergy effectively justifies our simplified treatment of the host gravitational potential from a kinematic perspective, while robustly validating the reliability of the Milky Way potential model and the initial orbital parameters adopted in this work. Based on these findings, we will discuss in detail in Section 4.2 how to utilize these simulation results as criteria to identify potential Whiting~1 member star candidates embedded within the Sgr background stellar stream.

In contrast, the simulated tidal structures for Eridanus (Figure \ref{fig:streamEr}) exhibit certain deviations from actual observations. 
We speculate that the cause of this discrepancy lies in the relatively limited dynamical studies and fundamental observational parameters (such as distance) for Eridanus, resulting in less-than-ideal constraints on its initial orbit. Furthermore, since Eridanus is located deep within the outer halo at a galactocentric distance of approximately 85 kpc, its orbital dynamical evolution is heavily influenced by the gravitational potential of the Milky Way's dark matter halo. Minor variations in the density profile of the dark matter halo (e.g., scale radius and triaxial shape) can significantly alter the phase-space trajectories of tidal tails over long-term orbital integrations, which serves as a more core systematic factor limiting our precise reconstruction of the historical orbit of the Eridanus. \citep{lawSagittariusDwarfGalaxy2010,bland-hawthornGalaxyContextStructural2016,vasiliev2021gaia}.  

\section{Discussion} \label{sec:highlight}

\subsection{Surface density}

Previous studies have indicated that the initial densities of GCs are typically much higher than their present-day observational values\citep{webb2015back,dickson2026fast}. Our dynamical simulations of Eridanus and Whiting 1 further corroborate this, revealing that the initial surface densities of these two clusters were 2 to 3 orders of magnitude greater than they are today. This stark contrast strongly suggests that they have undergone severe mass loss during their long-term dynamical evolution.

\begin{figure}[ht]
\centering
\includegraphics[width=\linewidth]{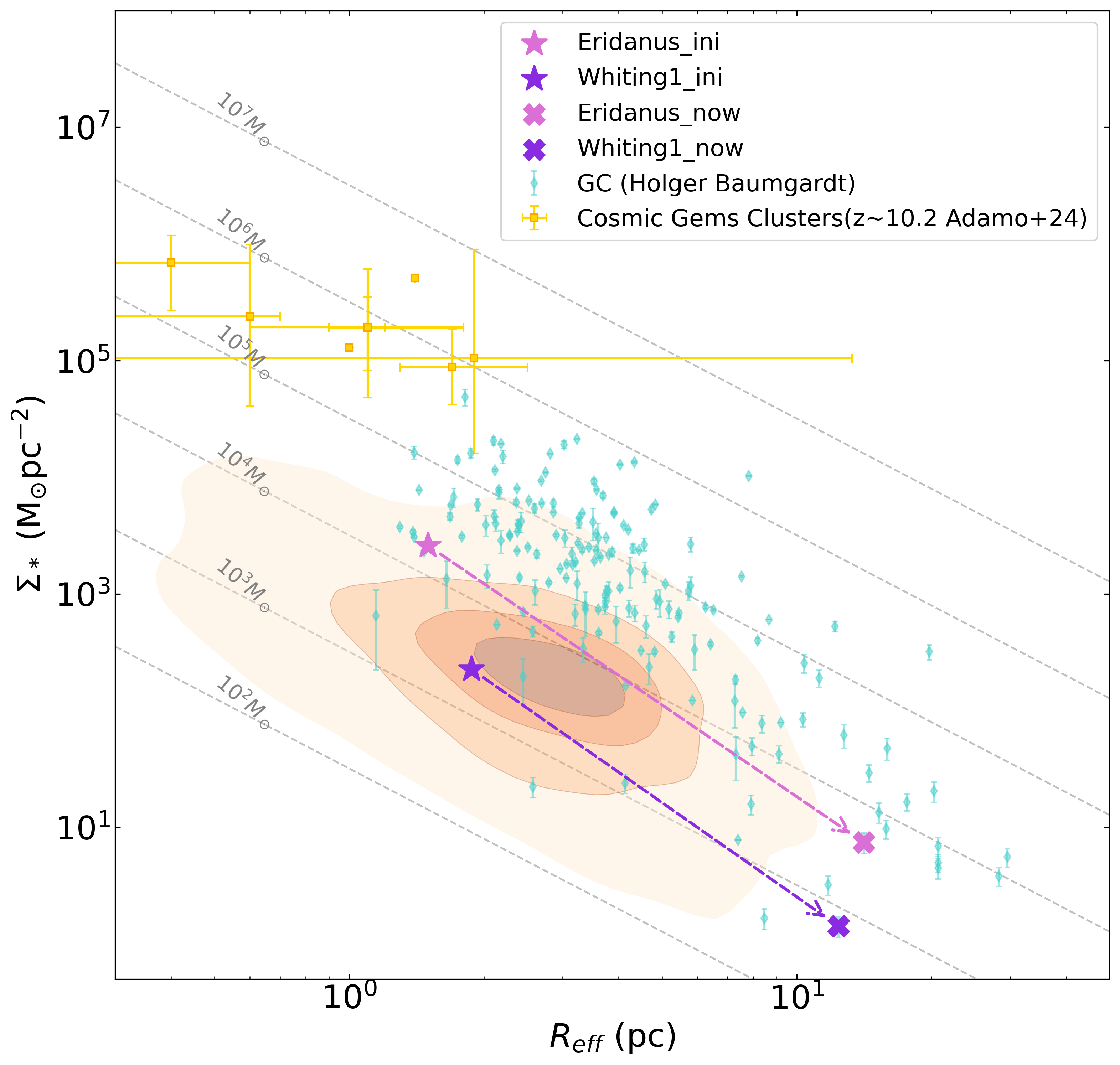} 

\caption{Distribution of star clusters in the $R_{eff}-\Sigma$ parameter space. Magenta and purple markers denote Eridanus and Whiting~1, respectively (stars: initial states; crosses: present-day). Comparison samples include Galactic GCs (cyan diamonds), high-redshift Cosmic Gems (blue squares; \citealt{adamoBoundStarClusters2024c}), and YMCs from the LEGUS survey (orange shading; \citealt{brown2021radii}). Grey diagonal lines indicate constant mass. $R_{eff}$ denotes the projected half-light radius ($r_{h,l}$) for Galactic GCs , and the projected half-mass radius ($r_{h}$) for all simulated models  and the 'Bounded GC' samples.}
\label{fig:den}
\end{figure}

This evolutionary trend suggests that if the extant population of GCs universally experienced an initial enhancement in their surface mass density ($\Sigma$) by two to three orders of magnitude, accompanied by a corresponding reduction in their half-mass radii ($r_h$), their primordial positions within the parameter space ($\Sigma$ vs. $R_{h,m}$) would shift collectively toward the upper left of Figure \ref{fig:den}.
This leads to an intriguing conjecture: the backward-modeled initial distribution zone aligns remarkably well with the recently discovered high-redshift ($z \sim 10$) compact star cluster candidates observed by JWST (e.g., the Cosmic Gems Cluster, denoted by blue squares in Figure \ref{fig:den}). This alignment hints at the possibility that the diffuse clusters currently residing in the outer halo may have shared a similar physical origin with these extremely compact objects from the Cosmic Dawn. However, given the limited sample size of our current work, this potential evolutionary scenario remains a preliminary hypothesis, which warrants systematic validation in future studies utilizing large-sample numerical simulations embedded with more diverse initial conditions.

Additionally, it is worth noting that within the Local Group and extending to spatial scales of several megaparsecs (Mpc), observations have confirmed the existence of nearly a hundred Young Massive Clusters (YMCs,\citealt{portegies2010young}). These systems universally exhibit spherical configurations, highly gravitationally bound states, and extreme compactness, with total masses comparable to or even exceeding those of typical ancient GCs. A profound evolutionary connection between these two populations is widely accepted within the astrophysical community; based on current observational and theoretical evidence, the initial formation mechanism of at least the metal-rich GC subpopulation is highly likely consistent with that of the YMCs observed in the local Universe.

Steered by this evolutionary framework, this study reveals that the initial parameters of Whiting~1 fall precisely within the parameter space of present-day YMCs. The absence of MPs in Whiting~1 \citep{huang2024driving} perfectly resonates with the mainstream observational consensus established for YMCs \citep{portegies2010young}. In this sense, modern YMCs may represent the younger evolutionary counterparts of old, MP-free star clusters like Whiting~1. In contrast, the initial parameters of Eridanus lie at the high-density extreme of the YMC distribution. Under a comparable surface density regime, the presence of MPs within Eridanus can be naturally explained by its significantly lower metallicity ($\text{[Fe/H]} \sim -1.45$) relative to modern YMCs. This striking contrast robustly substantiates our core physical argument that the driving mechanism behind MP formation is heavily governed by the dual constraint of metallicity and initial cluster compactness \citep{huang2024driving}.

It is worth noting that primordial binary systems were not included in these simulations. According to the simulation studies by \citealt{wang2024influence}, the inclusion of binaries has a negligible impact on the macroscopic structural evolution of the clusters investigated in this work.

\subsection{Whiting~1 Stream candidates}

Studies have shown that the Whiting~1 cluster originated within the Sagittarius dwarf spheroidal galaxy and was subsequently tidally-stripped during its accretion by the Milky Way. Previously, \citet{nie2019detecting} and \citet{nie2022searching} identified faint tidal extensions of Whiting~1. They both align with the cluster's orbit and the Sagittarius trailing stream. Our {\sc petar} simulations successfully reproduce the observed track of Whiting~1's stream (Figure \ref{fig:streamw1}), which is fully embedded in the Sgr stellar stream \citep{ramosFull5DCharacterisation2020, antojaAllskyPropermotionMap2020}. 

Identifying extratidal members of Whiting~1 from the Sgr stellar stream is challenging, but necessary to fully constrain the cluster dissolution. 
Using the validated simulation results as parameter filtering conditions, we identified potential extratidal members of Whiting 1 in the Gaia dataset. We established the following parameter boundaries to distinguish candidates from the field background:

\begin{enumerate}
    \item{$0^\circ<$ R.A. $<80^\circ$}
    \item{$-20^\circ<$ Dec. $<20^\circ$}
    \item $-2.234 < \mu_\alpha cos\delta < 1.414 \ { mas 
    \cdot yr}^{-1}$
    \item$-4.248 < \mu_\delta< 0.084  \ mas \cdot yr^{-1}$
\end{enumerate}

\begin{deluxetable*}{lcccccccccccccc}
\tabletypesize{\scriptsize}
\setlength{\tabcolsep}{3pt} 
\tablewidth{0pt} 
\tablecaption{Parameters of Selected Whiting~1 Stream Candidates \label{tab:full_params}}
\tablehead{
\colhead{ID} & \colhead{RA} & \colhead{Dec} & \colhead{RV} & \colhead{$\mu_{\alpha}\cos\delta$} & \colhead{$\mu_{\delta}$} & \colhead{$R_\odot$} & \colhead{$T_{\rm eff}$} & \colhead{$\log g$} & \colhead{[Fe/H]} &\colhead{[Fe/H]err} & \colhead{[C/Fe]} & \colhead{[N/Fe]} & \colhead{[Mg/Fe]} & \colhead{[Ca/Fe]} \\
\colhead{} & \colhead{(deg)} & \colhead{(deg)} & \colhead{(km s$^{-1}$)} & \colhead{(mas yr$^{-1}$)} & \colhead{(mas yr$^{-1}$)} & \colhead{(kpc)} & \colhead{(K)} & \colhead{(dex)} &\colhead{(dex)} & \colhead{(dex)} & \colhead{(dex)} & \colhead{(dex)} & \colhead{(dex)} & \colhead{(dex)}
}
\startdata
C1 & 17.15493 & -9.13032 & -115.19 & -0.671 & -2.568 & 20.48 & 4026 & 1.01 & -0.635 &0.012 & -0.361 & -0.023 & -0.029 & 0.038 \\
C2 & 28.33502 & -3.46134 & -133.81 & -0.422 & -2.473 & 22.37 & 3983 & 0.89 & -0.573 & 0.016&-0.374 & -0.018 & -0.036 & -0.030 \\
C3 & 21.07357 & -6.84508 & -119.42 & -0.457 & -2.675 & 24.71 & 3973 & 0.70 & -0.584 & 0.016&-0.410 & 0.047 & -0.089 & 0.009 \\
C4 & 17.87042 & -8.40780 & -124.19 & -0.604 & -2.361 & 21.81 & 3966 & 0.89 & -0.666 &0.014& -0.379 & -0.009 & -0.034 & 0.011 \\
C5 & 27.86607 & -3.87337 & -127.61 & -0.520 & -2.378 & 20.68 & 3913 & 0.77 & -0.677 & 0.014&-0.323 & -0.040 & -0.058 & 0.050 \\
\enddata
\tablecomments{Astrometric and photometric data are retrieved from Gaia DR3 \citep{vallenari2023gaia}, and spectroscopic parameters are from APOGEE DR17 \citep{abdurro2022seventeenth}. ID corresponds to the candidates identified in this study.}
\end{deluxetable*}

\begin{figure*}[ht]
\centering
\includegraphics[scale=0.5]{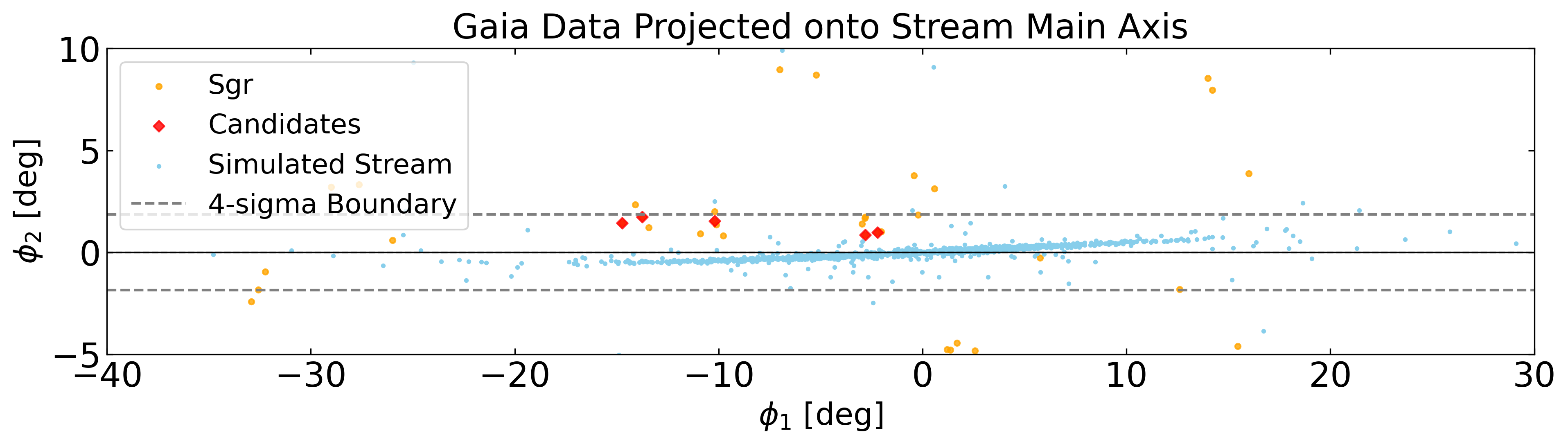}
\caption{Spatial distribution of the identified candidates in the stream-aligned coordinate system ($\phi_1, \phi_2$). The coordinate system has been transformed such that the main track of the Whiting~1 stream aligns with the $\phi_1$ axis. The cyan background points represent the member stars of the simulation, while the red diamonds denote our selected stream candidates. Orange points mark Sagittarius stream members with available chemical abundance information. The dashed lines represent the $3\sigma$ width of the simulated Whiting~1 stream track. 
}
\label{fig:phi}
\end{figure*}

To define the tidal tail boundaries, we transformed the sample into a stream-aligned coordinate system ($\phi_1, \phi_2$). We then calculated the standard deviation of the spatial spread perpendicular to the stream ($\sigma_{\phi_2}$) and retained only stars within $\pm 4 \sigma_{\phi_2}$ (Figure \ref{fig:phi}).

Next, we cross-matched candidates with APOGEE, selecting stars within a heliocentric distance of $20 - 40$ kpc (centered on Whiting~1, $d_\odot \approx 30$ kpc). 
We identified five RGB stars sharing similar dynamical and chemical features as Whiting~1, and thus they are strong candidates of Whiting~1 extratidal members (Table \ref{tab:full_params}).

\begin{figure*}[ht]
\centering
\includegraphics[width=\textwidth]{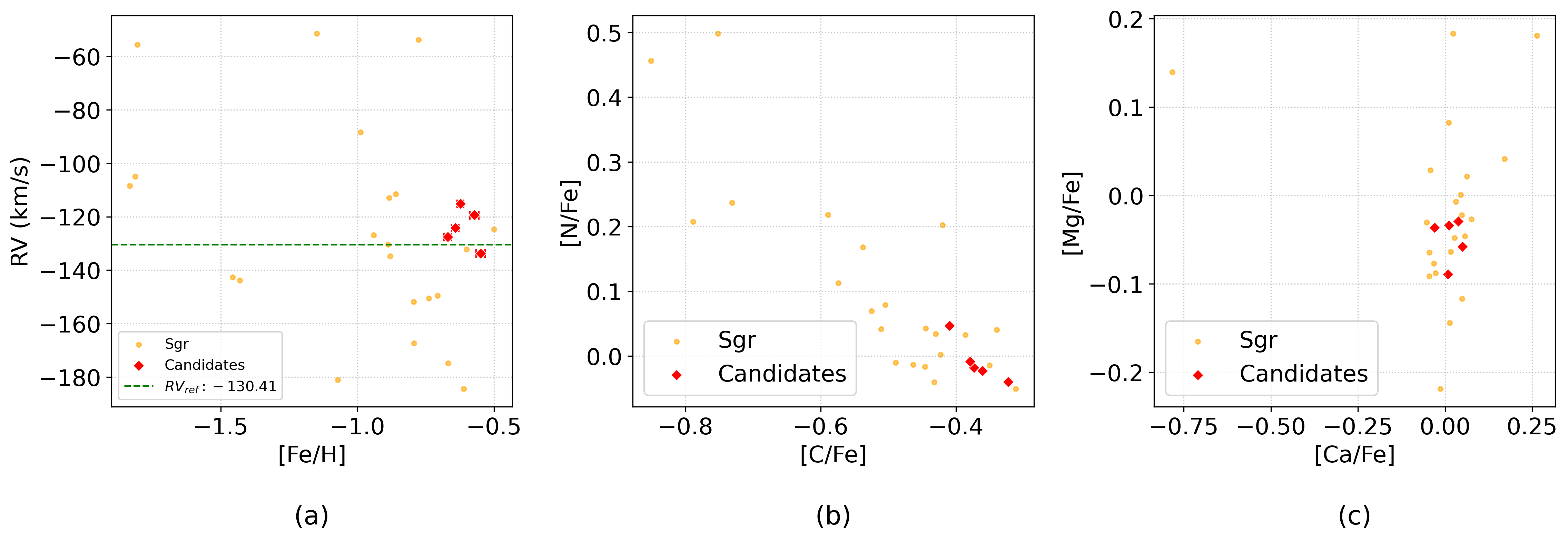}
\caption{Comparison of chemical abundances and kinematics between Whiting~1 extratidal candidates (red diamonds) and Sgr stream stars (orange circles).
(a) Radial velocity (RV) versus $[\mathrm{Fe}/\mathrm{H}]$. The red dashed line marks the typical RV of Whiting~1 ($-130.41~\mathrm{km~s^{-1}}$);
(b) $[\mathrm{C}/\mathrm{Fe}]$ versus $[\mathrm{N}/\mathrm{Fe}]$;
(c) [Mg/Fe] versus $[\mathrm{Ca}/\mathrm{Fe}]$.
}
\label{fig:cand}
\end{figure*}

\begin{enumerate}
    \item \textbf{Radial velocity and metallicity:} The five Whiting~1 extratidal candidates show similar RVs and metallicites (Figure \ref{fig:cand}a). Their RVs generally agree with the value of Whiting~1 (130 km/s, \citealt{vasiliev2021gaia}), with a scatter of $\sim 10$ km/s. This is larger than the uncertainties of APOGEE RV measurements ($< 0.4$ km/s), but it is not unexpected given that they may slightly change RVs when escaping the cluster. Meanwhile,
    the metallicities ([Fe/H]) of these candidates are highly consistent with the literature value for Whiting~1 ($-0.583$, \citealt{huang2024driving}).
    \item \textbf{Chemical abundance:} The C and N abundance distributions of these candidates fall within the parameter range empirically measured by \citet{huang2024driving} for Whiting~1 (Figure \ref{fig:cand}b). 
    Notably, in contrast to the Sgr stellar stream stars, the Whiting~1 extratidal candidates exhibit significantly stronger clustering in C, N, Mg and Ca (Figure \ref{fig:cand}b \& c), indicating a high degree of internal chemical homogeneity.
\end{enumerate}


The presence of candidate stars located beyond the $3\sigma$ spatial threshold of the simulated stream can be attributed to two primary physical factors. First, the actual Milky Way gravitational potential may deviate from the static model adopted in our N-body simulations.Given that the spatial offsets of these candidates are only of the order of a few degrees, a slight change in the potential model could naturally yield a shifted tidal tail that better encompasses these observed positions\citep{lawSagittariusDwarfGalaxy2010,bland-hawthornGalaxyContextStructural2016,vasiliev2021gaia}. Furthermore, our current simulations do not incorporate primordial binaries. In realistic star cluster environments, strong dynamical interactions involving binary stars can efficiently eject stars to significantly greater distances from the main stream track\citep{fujiiOriginOBRunaway2011}. The absence of this dynamical ejection mechanism in our models could thus account for the underprediction of such widely dispersed members. 
Notably,, due to observational selection effects, the identified stream candidates are predominantly high-luminosity stars with low $\log g$. According to the IMF, a substantial population of faint stars below the detection limit must exist within the stream, which leads to an underestimation of the cluster's mass loss.

\section{Conclusion} \label{sec:cite}

In this study, we employed the high-precision $N$-body code {\sc petar} to perform direct simulations of two representative low-mass Galactic GCs, Eridanus and Whiting~1. Our goal is to reconstruct their dynamical evolutionary histories and constrain their physical conditions at birth. This is part of our ongoing research project, ``Scrutinizing {\bf GA}laxy-{\bf ST}a{\bf R} cluster coevoluti{\bf ON} with chem{\bf O}dyna{\bf MI}cs ({\bf GASTRONOMI})'', which leverages multi-wavelength photometric and spectroscopic data to unravel the coevolutionary relationships between the MW, its satellite dwarf galaxies, and their star clusters.
By matching the simulation results with the observed present-day number density profiles, we determined the best-fit $r_{h,0}$ to be $2.0\pm1.0$ pc for Eridanus and $2.5\pm0.5$ pc for Whiting~1. These values correspond to initial surface densities of $1458.02\ M_\odot \cdot \text{pc}^{-2}$ and $127.33\ M_\odot \cdot \text{pc}^{-2}$, respectively, representing an enhancement of 2–3 orders of magnitude compared to their present-day states.

If this trend --- higher initial density and smaller radii at birth, is generalized to the Galactic GC population as a whole, the inferred initial conditions become consistent with the parameters of compact star clusters observed in high-redshift galaxies. Despite a surface density enhancement of two orders of magnitude, the initial state of Whiting~1 resembles that of YMCs in the local Universe. The absence of MPs in both Whiting~1 and YMCs suggests that MP formation generally requires higher initial surface density at a given radius, similar to that found in most present-day dense GCs.

Finally, using the simulated tidal stream morphology as a template for mining large-scale survey data, we successfully identified five high-probability extra-tidal candidates associated with Whiting~1. These candidates exhibit strong clustering in their kinematic and chemical abundances, with estimated values close to those of confirmed Whiting~1 members. With the advent of ongoing high-resolution spectroscopic surveys (e.g., 4MOST, WEAVE) reaching fainter stars, our simulations will serve as the ideal tool to uncover high-probability extra-tidal candidates of star clusters. Fully mapping these escaped stars is crucial for revealing the true initial conditions of clusters and their subsequent dissolution observationally.

In future work, we plan to extend these dynamical simulations to other Galactic GCs. By synthesizing our current observational and simulated results with future systematic studies, we hope to establish stronger physical connections between local Galactic GCs and early clusters at high redshifts, ultimately shedding light on the origin and formation mechanisms of GCs.

\begin{acknowledgments}
R. Huang and B. Tang gratefully acknowledge support from the National Natural Science Foundation of China through grants NOs. 12233013 and 12473035, China Manned Space Project under grant NO. CMS-CSST-2025-A13 and CMS-CSST-2021-A08, the Fundamental Research Funds for the Central Universities, Sun Yat-sen University (24qnpy121).

ChatGPT and Google Gemini were utilized solely for English language polishing during the drafting of this manuscript.
\end{acknowledgments}

\appendix
\section{Initial Radius Constraints and Mass Function Comparison}
\label{app:mass_function}

To validate the initial structural parameters of Eridanus, we compared the evolved present-day stellar mass functions ($dN/dm$) from our $N$-body simulations initialized with different half-mass radii ($r_{h,0} = 1.5$, $2.0$, and $3.0\,\text{pc}$) against HST star counts.
\begin{figure*}[htbp]
    \centering
    \begin{subfigure}[b]{0.32\textwidth}
        \centering
        \includegraphics[width=\textwidth]{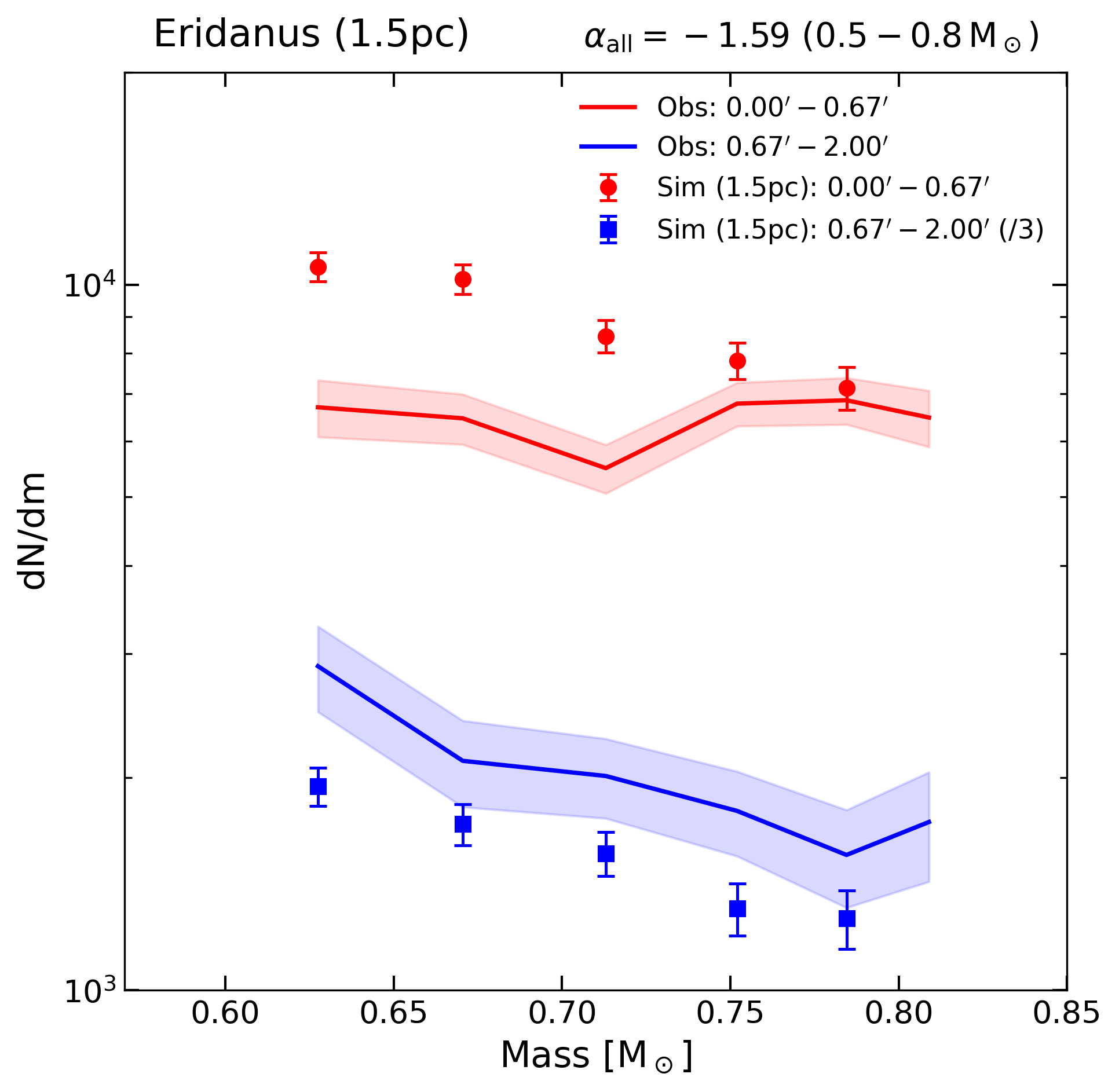}
        \caption{$r_{h,0} = 1.5\,\text{pc}$}
        \label{fig:eridanus_mf_1.5pc}
    \end{subfigure}
    \hfill
    \begin{subfigure}[b]{0.32\textwidth}
        \centering
        \includegraphics[width=\textwidth]{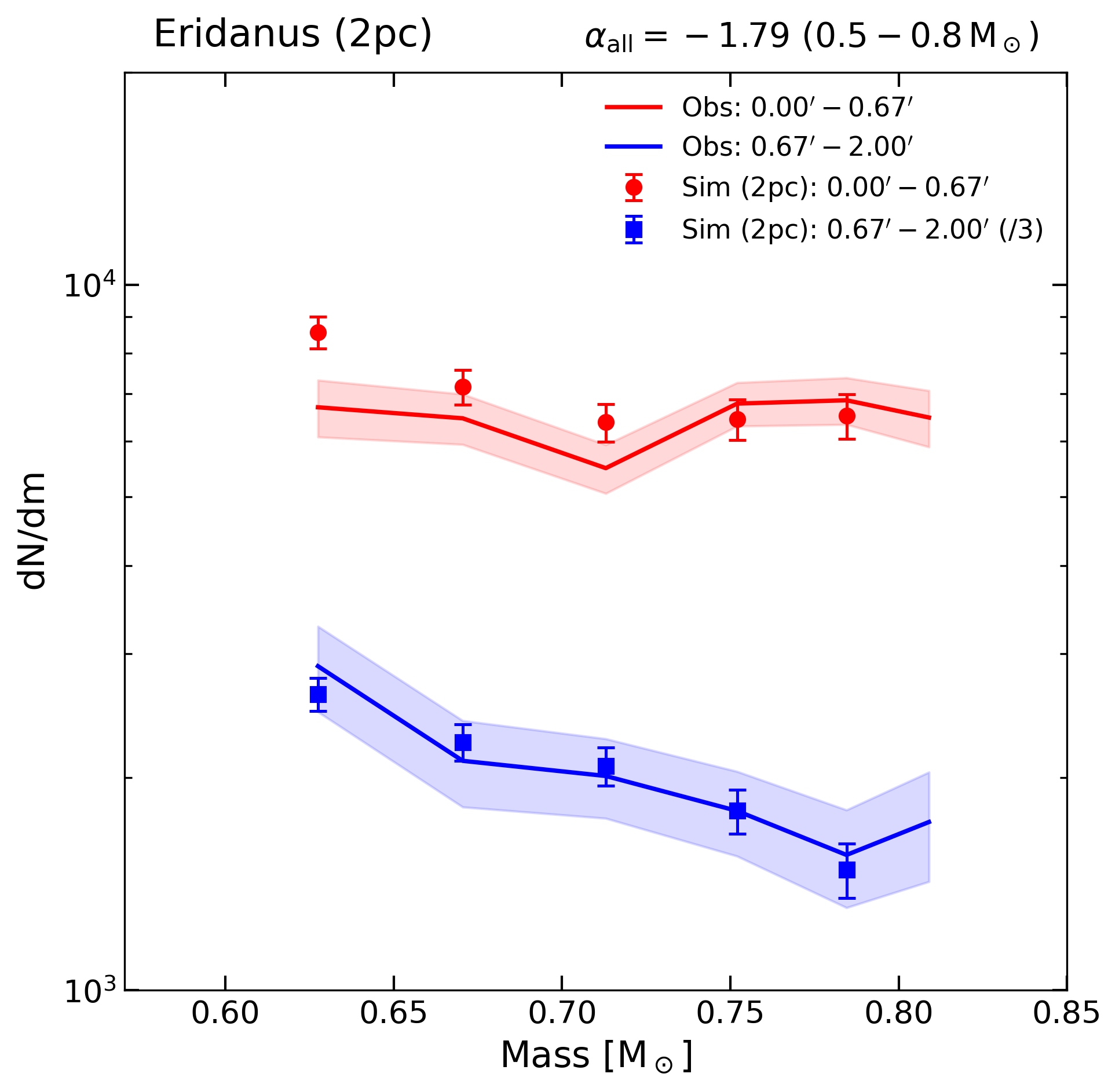}
        \caption{$r_{h,0} = 2.0\,\text{pc}$ (Best fit)}
        \label{fig:eridanus_mf_2pc}
    \end{subfigure}
    \hfill
    \begin{subfigure}[b]{0.32\textwidth}
        \centering
        \includegraphics[width=\textwidth]{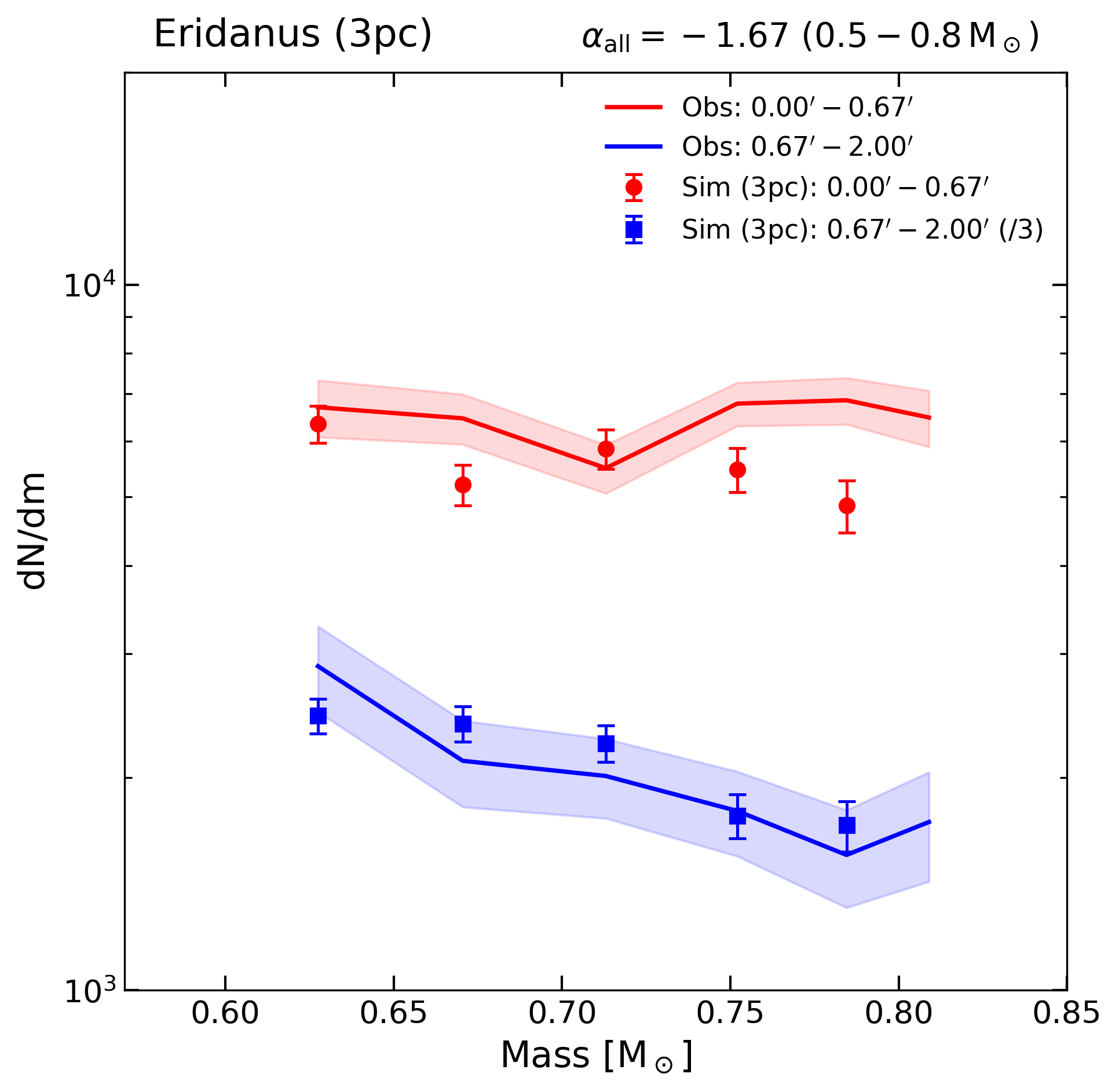}
        \caption{$r_{h,0} = 3.0\,\text{pc}$}
        \label{fig:eridanus_mf_3pc}
    \end{subfigure}
    \caption{Radial stellar mass functions ($dN/dm$) of Eridanus comparing HST photometric observations from \citet{baumgardt2023evidence} (solid lines with shaded uncertainty bands) with $N$-body simulations initialized with different half-mass radii: (a) $r_{h,0} = 1.5\,\text{pc}$, (b) $r_{h,0} = 2.0\,\text{pc}$, and (c) $r_{h,0} = 3.0\,\text{pc}$ (points with Poisson error bars). Red markers and lines correspond to the inner cluster region ($0.00'-0.67'$), while blue markers and lines represent the outer region ($0.67'-2.00'$). For visual clarity, the outer radial profiles in all panels are scaled downward by a factor of 3. The $r_{h,0} = 2.0\,\text{pc}$ model provides the best fit to the observed mass segregation profile across the entire mass range.}
    \label{fig:eridanus_mf_comparison}
\end{figure*}

Figure~\ref{fig:eridanus_mf_comparison} compares the evolved present-day stellar mass functions ($dN/dm$) for three different initial half-mass radii ($r_{h,0} = 1.5$, $2.0$, and $3.0\,\text{pc}$) against the HST star counts. 

As shown in panel (b), the $r_{h,0} = 2.0\,\text{pc}$ model yields the best overall match to both the normalization and the mass segregation slope between the core ($0.00'-0.67'$) and outer halo ($0.67'-2.00'$). In contrast, the more compact configuration with $r_{h,0} = 1.5\,\text{pc}$ (panel a) suffers from excessive tidal stripping, which overdepletes low-mass stars, while the more extended model with $r_{h,0} = 3.0\,\text{pc}$ (panel c) remains dynamically under-evolved and fails to reproduce the observed mass segregation gradient. Since the HST observational data for Whiting~1 remain proprietary, no corresponding mass function comparison is presented here.

\bibliography{sample701}{}
\bibliographystyle{aasjournalv7}



\end{document}